# Quenching Current by Flux-Flow Instability in Iron-Chalcogenides Thin Films

A. Leo, G. Grimaldi, A. Guarino, F. Avitabile, P. Marra, R. Citro, V. Braccini, E. Bellingeri, C. Ferdeghini, S. Pace, A. Nigro

***Abstract*—The stability against quench is one of the main issue to be pursued in a superconducting material which should be able to perform at very high levels of current densities. Here we focus on the connection between the critical current $I_c$ and the quenching current $I^*$ associated to the so-called Flux-Flow Instability phenomenon, which sets-in as an abrupt transition from the flux flow state to the normal state. To this purpose, we analyze several current-voltage characteristics of three types of Iron-Based (IBS) thin films, acquired at different temperature and applied magnetic field values. For these samples, we discuss the impact of a possible coexistence of intrinsic electronic mechanisms and extrinsic thermal effects on the quenching current dependence upon the applied magnetic field. The differences between the quenching current and the critical current are reported also in the case of predominant intrinsic mechanisms. Carrying out a comparison with the HTS case, we suggest which material can be the best trade-off between maximum operating temperature, higher upper critical field and stability under high current bias.**

## I. Introduction

THE CHOICE OF a particular superconducting material over another for a specific application is made taking into account several parameters, such as the critical temperature $T_c$, the upper critical field $B_{c2}$ and the critical current $I_c$.

A high critical temperature allows to use cheaper cryogenics, as in the case of High Temperature Superconductors (HTSs), which can operate, for example, in liquid Nitrogen. The use of a superconducting material with a high upper critical field is mandatory in those devices which are intended to generate or operate in high magnetic fields. In this case, the Iron Based Superconductors (IBSs) are promising materials [1].

The critical current density determines the maximum operating current of the device, which is typically set to 2/3 of $I_c$ in order to minimize the impact of fluctuations. Indeed, a variation in the operating temperature or in the applied magnetic field can depress the critical current and drive the superconducting material into a dissipative state above $I_c$, commonly referred for type-II superconductors as the flux-flow state, since in this case the dissipation arises from the motion of the Abrikosov vortices. Once in the dissipative state, the subsequent self-heating can trigger a thermal runaway leading to the quench to the normal state. Usually, this transition is supposed to occur gradually, but often this does not happen. Indeed, it is well known that almost all superconducting materials, including HTSs and IBSs, can be affected by the presence of the Flux-Flow Instability (FFI) phenomenon [2]–[12]. This phenomenon is associated with an instability of the vortices at large driving forces [2], [13]-[15]. In the current biased current-voltage characteristics (CVCs), the typical signature of the FFI is an abrupt voltage jump associated with the transition from the flux-flow state to the normal conduction. Thus, this phenomenon can be relevant for applications, since it can strongly compromise the high-current-carrying capability of any superconducting device. The presence of FFI in a superconducting sample is related to different material properties, among them we can mention the pinning due to natural or artificial defects, or the sample geometry [16]–[23]. In particular, the FFI can be observed in samples whose dimensions are comparable to those of the filaments in superconducting cables.

In a current biased CVC two critical parameters identify the instability point, i.e. the point at which the FFI jump is observed: the critical voltage $V^*$ and the quenching current $I^*$ (also referred in the literature as instability or supercritical current). In the range between $I_c$ and $I^*$ the material is still in the superconducting phase, but dissipation due to moving vortices is present. The relation between this two critical current values is a high relevant topic to the potential high field applications of superconducting materials, but less well studied in the community, with few works on HTSs to our knowledge [24], [25]. In this work, we investigate the relation between the critical current and the quenching current in Iron Based Superconductors, in particular in the Fe-chalcogenide compound Fe(Se,Te). We focus our analysis on the difference between $I_c$ and $I^*$, namely $I^*$-$I_c$, which can be seen as a safe range before the complete quench of the superconductor. We show that a crossover between two different behaviors as a function of the field is present, which is not comparable to those observed in HTSs, but it looks like to those found in Low

The research leading to these results has received funding from the PON Ricerca e Competitività 2007-2013 under grant agreement PON NAFASSY, PONa3_00007. (All authors contributed equally to this work.) (Corresponding author: Antonio Leo.)

A. Leo, A. Guarino, R. Citro, A. Nigro, and S. Pace are with Physics Department ‘E. R. Caianiello’, Salerno University, via Giovanni Paolo II, 132, Stecca 9, I-84084 Fisciano (SA), Italy. They are also with CNR-SPIN Salerno, via Giovanni Paolo II, 132, Stecca 9, I-84084 Fisciano (SA), Italy, e-mail: antonio.leo@fisica.unisa.it; guarino@fisica.unisa.it; citro@fisica.unisa.it, nigro@fisica.unisa.it; pace@fisica.unisa.it.

G. Grimaldi and P. Marra is with CNR-SPIN Salerno, via Giovanni Paolo II, 132, Stecca 9, I-84084 Fisciano (SA), Italy, e-mail: gaia.grimaldi@spin.cnr.it; pasquale.marra@spin.cnr.it.

F. Avitabile is with Physics Department ‘E. R. Caianiello’, Salerno University, via Giovanni Paolo II, 132, Stecca 9, I-84084 Fisciano (SA), Italy, e-mail: favitabile@unisa.it.

E. Bellingeri, V. Braccini, and C. Ferdeghini are with CNR-SPIN Genova, corso Perrone 24, I-16152 Genova, Italy, e-mail: emilio.bellingeri@spin.cnr.it; valeria.braccini@spin.cnr.it; carlo.ferdeghini@spin.cnr.it.

Temperature Superconductors (LTSs).

## II. Experimental Details

### *A. Measurement setup*

The measurement setup is based on a Cryogenic Ltd. cryogen free cryostat equipped with an integrated cryogen-free variable-temperature insert operating in the range 1.6–300 K and a superconducting magnet able to generate a field up to 16 T. In this system, the metallic sample holder is cooled by a continuous Helium gas flow and the temperature stability is within 0.01 K. Sample temperature is measured via a LakeShore Temperature Controller model 350 connected to a LakeShore Cernox sensor model CX-1030-SD-1.4L mounted on the same metallic block used as sample holder.

Resistivity measurements as a function of the temperature in different magnetic fields applied perpendicularly to the film ab plane have been performed with a standard 4-probe technique using a Keithley model 2430 as current source and a Keithley model 2182 as voltage meter. The critical temperature at zero applied magnetic field $T_c(0)$ and the upper critical field at zero temperature $B_{c2}(0)$ have been evaluated from these measurements. The $T_c(0)$ is defined as the temperature value in absence of a magnetic field at which the onset of the superconducting transition is observed. A rough estimation of $B_{c2}(0)$ can be obtained from the upper critical field curve as a function of the temperature via the Werthamer-Helfand-Hohenberg (WHH) formula with a Maki parameter of about 2.5 [26]. We note that the values of $B_{c2}(0)$ obtained with this approach are usually under-estimated and that more precise values can be obtained taking into account the effects of orbital and paramagnetic pair breaking as well as of FFLO instability in multi-band superconductors [27].

Current-Voltage Characteristics (CVCs) have been performed at different temperatures and external magnetic field values, with the magnetic field applied perpendicularly to the film ab plane. A pulsed current 4-probe technique has been used, with the Keithley model 2430 used both as current source and voltage meter. In our CVC measurement, each current pulse has a rectangular shape with a power-on time (or Pulse Width, PW) equal to 2.5 ms; the time separation between each pulse (or Pulse Delay, PD) is set to 1 s in order to allow complete recover of the sample temperature to the Helium flow temperature. As a consequence, no thermal hysteresis is observed in the acquired curves. The sample holder temperature $T$ is monitored during the whole CVC acquisition; $T$ values are acquired just before each current pulse. The critical current $I_C$ is evaluated from CVC by a standard 1 μV/cm criterion.

### *B. Samples description and properties*

The data here reported are related to microbridges obtained from Fe(Se,Te) thin films on $CaF_2$ (001) oriented substrate by standard UV photolithography and Ar ion-milling etching. The films have been fabricated by Pulsed Laser Deposition (PLD) starting from a target with the nominal composition $FeSe_{0.5}Te_{0.5}$. Three types of samples have been analyzed.

Sample named W1K is a wide bridge of width $W$ of about 1 mm, length (considered as the distance between voltage tips) $L$ of about 10 mm and thickness $S$ of about 100 nm. The typical $T_c(0)$ for this type of samples is 13.5 K, while $B_{c2}(0)$ is about 31.1 T. Samples W20 and W10 belong to an optimized second generation of thin films. The microbridge geometry is defined as $W$ = 20 μm, $L$ = 65 μm and $S$ = 120 nm for sample W20 and $W$ = 10 μm, $L$ = 50 μm and $S$ = 150 nm for sample W10. The $T_c(0)$ values are 20.5 K for sample W20 and 18.9 K for sample W10, while $B_{c2}(0)$ are respectively 41.8 T and 38.7 T. These data are summarized in Table I, while more information about sample fabrication and their structural and pinning properties can be found elsewhere [28], [29].

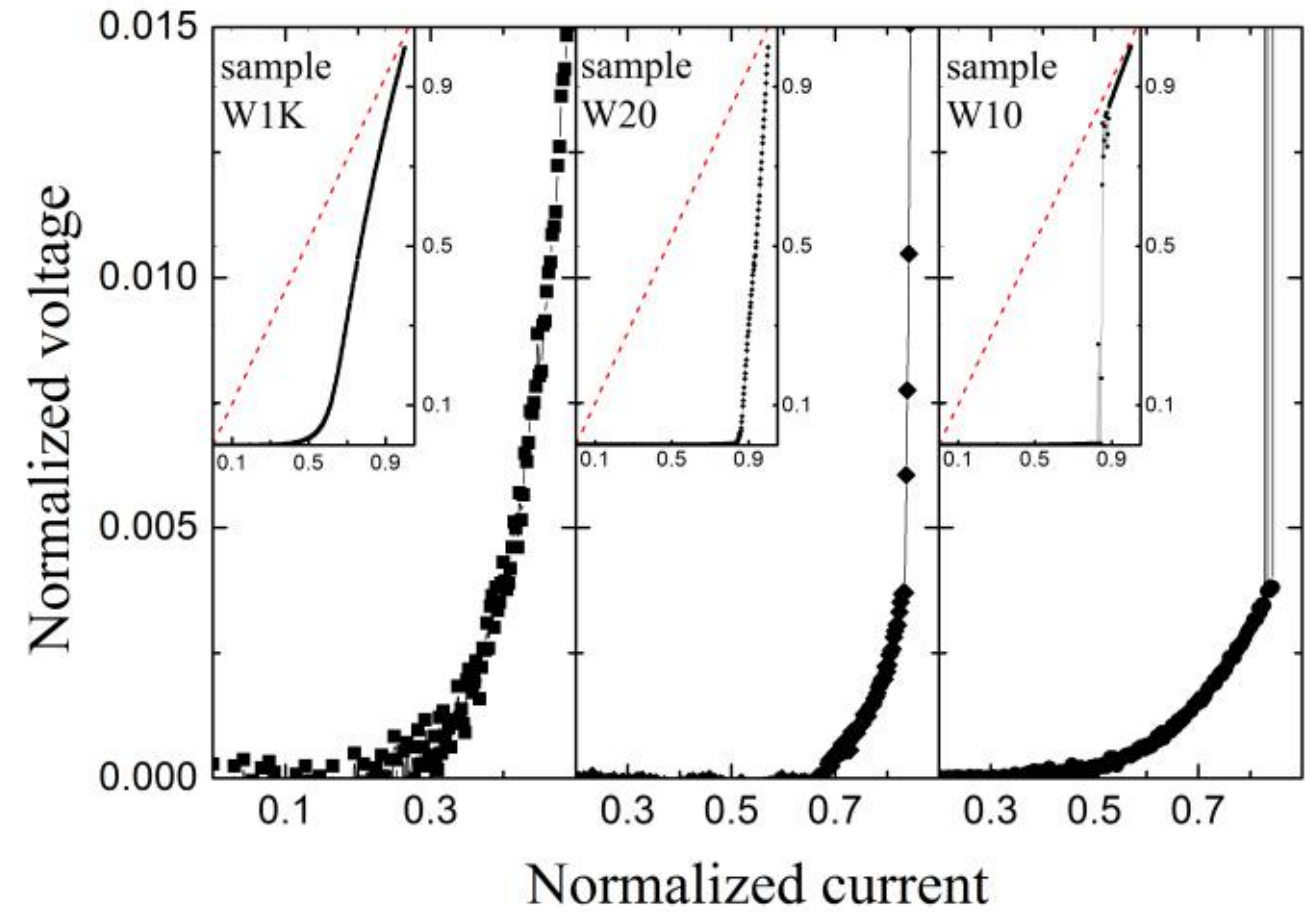

Fig. 1. Current-voltage characteristics of the three analyzed samples in the low voltage region. The current values are normalized to the maximum value of the current feeded to the sample, i.e. 30 mA for sample W1K, 19 mA for sample W20 and 3 mA for sample W10. The voltage values are normalized to the maximum measured value of the voltage, i.e. 0.8817 V for sample W1K, 1.39284 V for sample W20 and 0.93603 V for sample W10. The reduced temperature $t = T/T_C$ is 0.69 for sample W1K, 0.72 for sample W20 and 0.63 for sample W10. The field is 0.5 T for all three samples. (Insets) The three curves in the full range; the red dashed line is the normal resistivity line.

TABLE I
RELEVANT INFORMATION ON ANALYZED SAMPLES

| Sample | *W*, *L*, *S* | $T_C(0)$ | $H_{c2}(0)$ |
|---|---|---|---|
| W1K | 1 mm, 10 mm, 120 nm | 13.5 K | 31.1 T |
| W20 | 20 μm, 65 μm, 120 nm | 20.5 K | 41.8 T |
| W10 | 10 μm, 50 μm, 150 nm | 18.9 K | 38.7 T |

The table summarizes the relevant information on the analyzed samples. Here, $W$, $L$ and $S$ are the microbridge width, length and thickness respectively; $T_C(0)$ is the critical temperature at zero applied magnetic field; and $B_{c2}(0)$ is the upper critical field evaluated at zero temperature via the WHH formula with a Maki parameter of 2.5.

## III. Results and Discussion

### *A. Current-Voltage Characteristics and Flux Flow Instability*

The Flux-Flow Instability phenomenon can be triggered by intrinsic electronic [3], [13], [15] as well as by extrinsic thermal mechanisms [9], [10], [14]. In both cases, for the current biased current-voltage characteristics a voltage jump from the low dissipative regime up to the normal conduction state can be observed. The steepness of this transition strongly depends on which mechanism prevails.

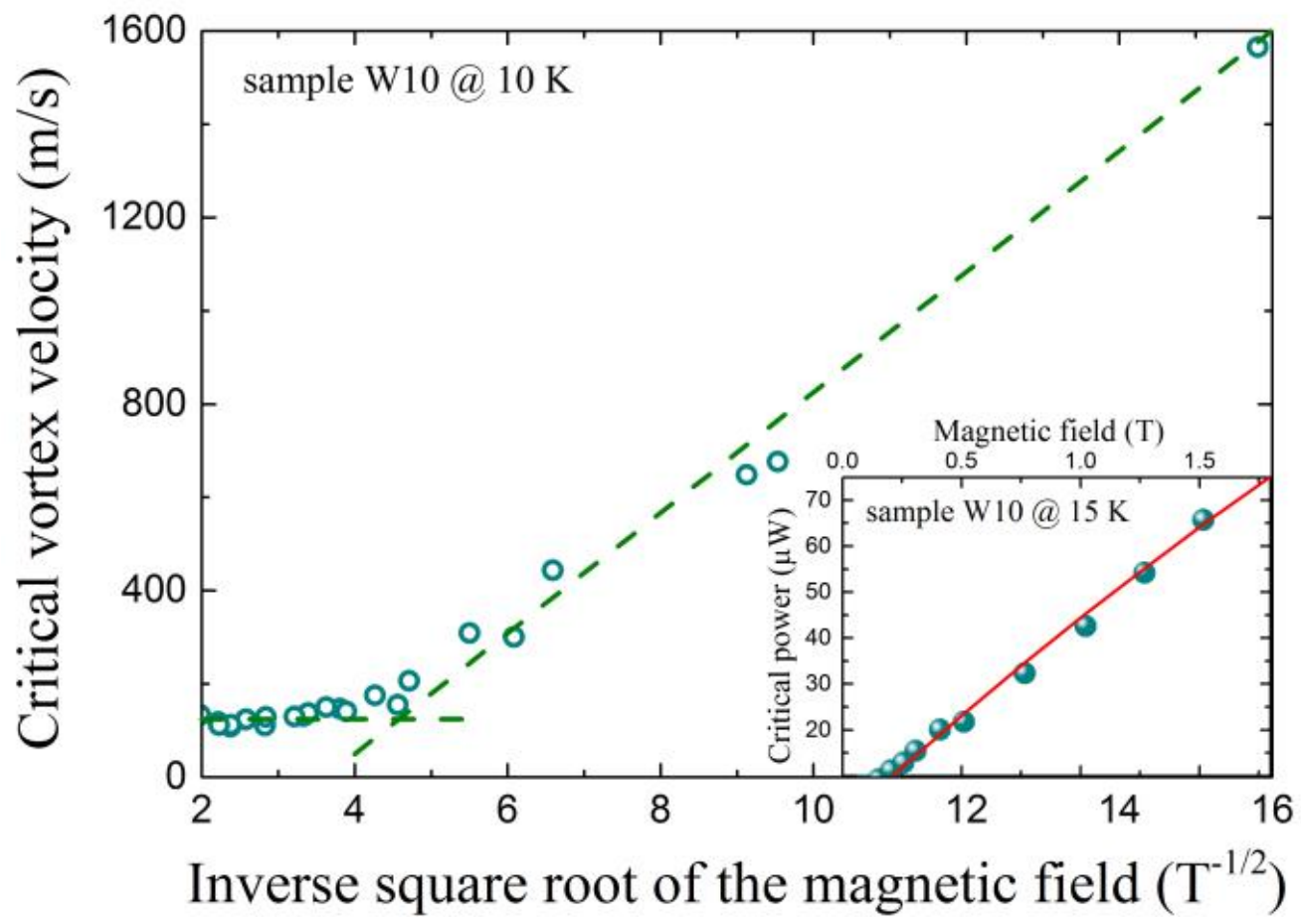

Fig. 2. Behavior of the critical vortex velocity as a function of the inverse of the square root of the applied magnetic field. The dashed lines are guides for the eye. (Inset) Curve of the critical power as a function of the applied magnetic field. The red solid line is the fitting curve resulting from the procedure described in the text.

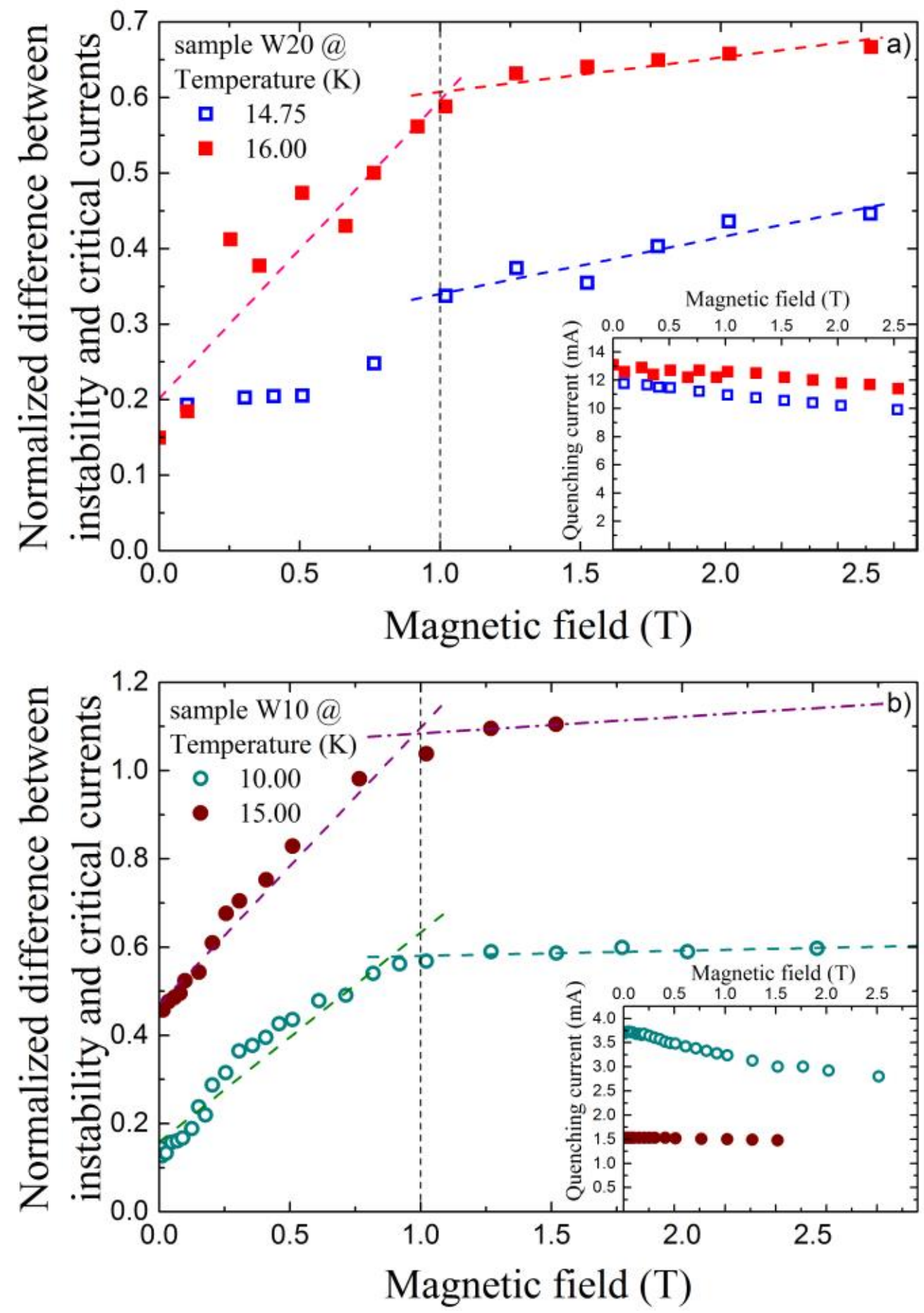

Fig. 3. Behavior of the relative value of the difference between the quenching and the critical current normalized to the critical current value at zero field at different temperatures for (a) sample W20 and (b) sample W10. Here, the data are normalized to the $I_c(0)$ values 11.7 mA at 14.75 K and 11.4 mA at 16.00 K for sample W20, and 3.24 mA at 10.00 K and 1.05 mA at 15.00 K for sample. The dashed lines are guides for the eye. (Insets) Curves of the quenching current as a function of the applied magnetic field for (a) sample W20 and (b) sample W10.

In Fig. 1, for each of the three considered samples, a typical CVC is shown. The CVC are acquired at similar reduced temperatures $t = T / T_c(0)$ and at the same applied magnetic field $B = 0.5$ T. Looking at the low voltage region, where the sudden voltage jump associated to the Flux-Flow Instability is usually observed, we note that in the case of sample W1K the signature of the phenomenon is not present. The lack of FFI can be explained by the wide geometry of the sample, which maximize the self-heating effects. Indeed, the heat-transfer coefficient $h_s$ between the sample and the substrate decreases as the bridge width is increased [22]. In wide bridges, the temperature increase due to the less efficient heat removal prevents the vortices from reaching the critical velocity. Although this observation strictly depends on the material under investigation as well as on the substrate and cooling environment [10], [23]. In the case of sample W20 the characteristic transition associated to FFI is observed. In this case, the presence of a smoothed jump is the result of the coexistence of thermal and electronic mechanisms [11], [12].

Sample W10 has been realized in order to obtain a predominance of electronic mechanisms by maximizing $h_s$. Thus, it is not surprising that the transition in this case is steeper than that of the sample W20. To support the hypothesis about the predominance of intrinsic mechanisms, we analyzed the magnetic field dependence of the critical vortex velocity $v^* = V^*/(L \cdot B)$. As shown in Fig. 2, we found that the $v^*$ behavior is in agreement with that expected in the case of a FFI triggered by electronic mechanisms, i.e. a $B^{-1/2}$ dependence at low field values followed by a rather constant value at higher fields [4], [5], [8], [9]. Moreover, we estimated the $B_T$ parameter introduced by Bezuglyj and Shklovskij (BS) [14], which separates the region where non-thermal intrinsic ($B << B_T$) or pure heating extrinsic mechanisms ($B >> B_T$) of the instability dominates. Following the BS approach, it is possible to estimate $B_T$ from the critical power $P^* = I^* \cdot V^*$ curve as a function of the applied magnetic field, since $P^* \propto (1 - t)$, where $t$ is a function of $B/B_T$ [14]. In the inset of Fig. 2 the $P^*(B)$ curve at 15.00 K is shown. From these data, we estimate $B_T = 20$ T, a value strictly above the considered field range.

### B. *Quenching Current vs. Critical Current*

The recognition of the Flux-Flow Instability in the current-voltage characteristics of samples W20 and W10 paves the way to the study of the relation between the quenching current and the critical current based on the possible coexistence of both thermal and electronic mechanisms, as well as on the predominant intrinsic mechanism.

In Fig. 3, the difference between the quenching current and the critical current normalized to the critical current value at zero field $\Delta i = (I^* - I_c)/I_c(0)$ is reported as a function of the applied magnetic field for both samples W20 and W10 at two different temperatures. In both cases, despite the fact that the increase of the temperature implies a reduction of the critical current, the values of $\Delta i$ are higher for the upper temperature value in the whole magnetic field range. This feature is a consequence of two concurrent effects. First, a higher stability in the current conduction at higher temperature, due to a weaker contribution from the extrinsic thermal mechanism to the FFI [9]. Second, the weaker influence of the pinning strength on the quenching current with respect to the influence on the critical current, regardless of the pinning mechanism [18]–[20].

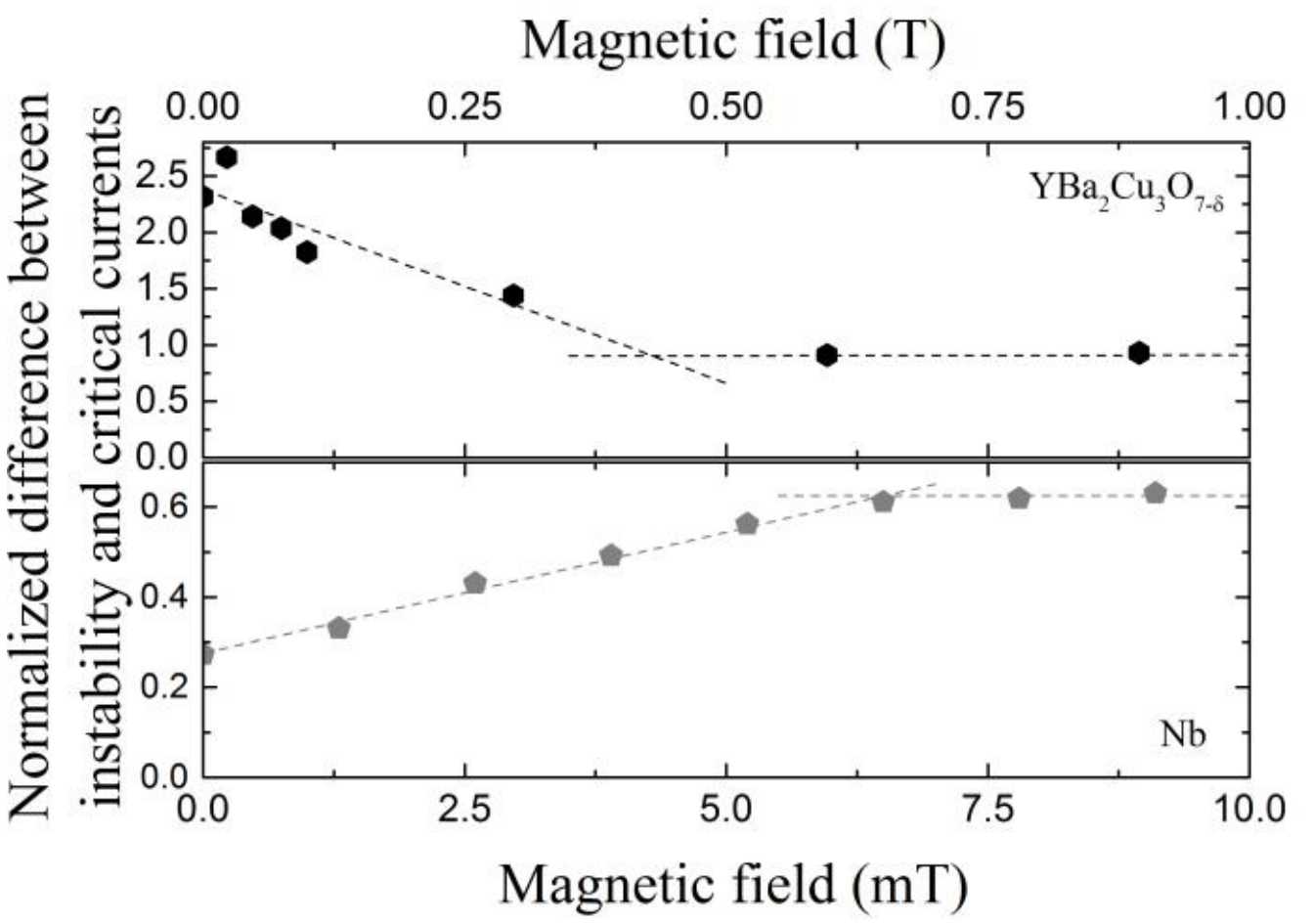

Fig. 4. Behavior of the relative value of the difference between the quenching and the critical current normalized to the critical current value at zero field and different temperatures for (upper panel) an HTS sample as by the data reported in [24] and (lower panel) a LTS sample as by the data reported in [7]. The dashed lines are guides for the eye.

Moreover, a crossover between an increasing $\Delta i$ and a quite constant behavior with the increasing applied magnetic field is observed in both samples. In particular, the crossover field value is around 1 T for all the curves. This behavior is still a consequence of the weaker dependence of the quenching current on the applied magnetic field with respect to the critical current. The substantial independence of $I^*$ on $B$ both in the case of a thermal origin of the instability and in the case of intrinsic electronic mechanisms triggering the FFI can be inferred by the data reported in the insets of Fig. 3. Indeed, the percentage variation of $I^*$ in the considered range is always less than 24%, with a minimum value of about 3% for sample W10 at 15 K. On the contrary, the percentage variation of $I_c$ in the same range is always more than 60%. In particular, in the case of intrinsic electronic mechamisns, the almost $B$-independent behavior of $I^*$ is the result of the influence of the peculiar pinning landscape [18], [19].

### *C. Comparison with other materials*

In Fig. 4 we show the $\Delta i$ curves as a function of the applied magnetic field as obtained by the data for HTS ($YBa_2Cu_3O_{7-\delta}$) reported by Doval et al. in [24] and for LTS (Nb) reported by Grimaldi et al. in [7]. We note that, in the case of HTS analyzed by Doval et al., the quenching current $I^*$ is associated with pure thermal mechanism, which Maza et al. claim to be the predominant mechanism for the FFI in HTS [10]. In this case, $I^*$ also results to be less affected than $I_c$ by the applied magnetic field, but the $\Delta i$ behavior is just the opposite of the one shown by our IBS samples. Indeed, in the HTS case, $\Delta i$ initially decreases as the field increases. On the contrary, in the LTS case the FFI has been proven to be triggered by pure electronic mechanisms [6], [7], [19]. In this case, we note a very similar behavior of $\Delta i(B)$ to those of our IBS. From these observations and considering the established high values of upper critical field and of its slope near the critical temperature, as well as the high values of the critical and the quenching currents in high magnetic fields, we can argue that Iron-Based Superconductors should be considered as High Field Superconductors with performance comparable to, or even better than those of High Temperature Superconductors.

## IV. Conclusion

In conclusion, quench features related to Flux-Flow Instabilities have been recognized in optimized Fe(Se,Te) thin films grown by Pulsed Laser Deposition on a $CaF_2$ (001) oriented substrate. In particular, the relation between the quenching current and the critical current has been analyzed both in the case of FFI ascribed to the coexistence of extrinsic thermal and intrinsic electronic mechanisms and in the case of predominant intrinsic electronic mechanisms.

On the basis of the results reported in the present work, we argue that the contribution of the intrinsic mechanisms to the FFI in IBS leads to a substantial independence of the quenching current from the intrinsic pinning influence. Moreover, we observe the presence of a crossover from an increasing difference between the quenching and the critical current to a quite constant value as the applied magnetic field is increased. This feature is observed for FFI driven by both thermal and electronic or by only electronic mechanisms, as in the presented cases of IBS or LTS. On the contrary, data taken from literature related to FFI driven by only thermal mechanisms, as in the case of HTS, show a quite opposite behavior. On these basis, it can be argued that the presence of a significant contribution from intrinsic mechanisms to the Flux-Flow Instabilities can be inferred by the analysis of the relation between the quenching current $I^*$ and the critical current $I_c$. Although, it cannot be excluded that the different behaviors are related to different superconducting properties between the materials, e.g. different gap structures, thus further investigation are needed.